\documentclass[amsmath,amssymb,aps,prl,twocolumn,showpacs]{revtex4}
\usepackage{epsfig,graphics,amsfonts,amssymb,amsmath,graphicx} 
\begin{document}

\title{``Swimming" versus ``swinging" effects in spacetime}
\author{Eduardo Gu\'eron}
\email[e-mail: ]{gueron@ime.unicamp.br}
\affiliation{Departamento de Matem\'atica Aplicada, 
    Instituto de Matem\'atica, Estat\'\i stica e 
    Computa\c c\~ao Cient\'\i fica\\
    Universidade de Campinas, 13083-970, 
    Campinas, SP, Brazil 
} 
\author{Cl\'ovis A. S. Maia}
\email[e-mail: ]{clovis@ift.unesp.br}
\author{George E. A. Matsas}
\email[e-mail: ]{matsas@ift.unesp.br} 
\affiliation{Instituto de F\'\i sica Te\'orica, 
             Universidade Estadual Paulista,
             01405-900, S\~ao Paulo, SP, Brazil }  

\begin{abstract}
Wisdom has recently unveiled a new {\em relativistic} effect, 
called ``spacetime swimming'', where quasi-rigid free bodies 
in curved spacetimes can ``speed up", ``slow down" or ``deviate"
their falls by performing {\em ``local"} cyclic shape deformations. 
We show here that for fast enough cycles this effect dominates over 
a {\em non-relativistic} related one, named here 
``space swinging'', where the fall is altered 
through {\em ``nonlocal"} cyclic deformations in Newtonian 
gravitational fields. We expect, therefore, to clarify the 
distinction between both effects leaving no room to controversy. 
Moreover, the leading contribution to the swimming effect 
predicted by Wisdom is enriched with a higher order term and
the whole result is generalized to be applicable in cases 
where the tripod is in large red-shift regions.
\end{abstract}

\maketitle

\begin{figure}[t]
\includegraphics[width=0.29\textwidth]{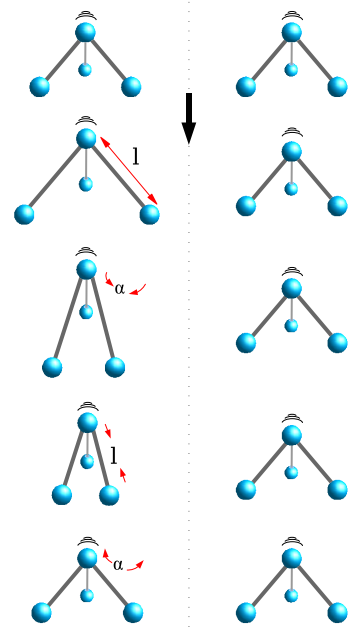}
\caption{Five snapshots of two tripods designed to have
legs with length $l$ and angle $\alpha$ with the radial 
axis, along which they fall down, are shown with and 
without cyclic deformations, respectively. The swimming  effect 
consists in realizing that {\em local} cyclic deformations 
lead, in general, to displacements of order $G/c^2$ in the 
quasi-rigid tripod trajectory when compared with the 
rigid one.}
\label{snapshots}
\end{figure}

Recently, Wisdom unveiled a new beautiful 
{\em relativistic effect}~\cite{W} (see also Ref.~\cite{BS}) 
denominated {\em spacetime swimming}, where quasi-rigid free 
bodies in curved spacetimes can ``speed up", ``slow down" 
or ``deviate" their falls by 
performing {\em ``local"} cyclic shape deformations 
(see Fig.~\ref{snapshots}). This is a full 
general-relativistic geometrical phase effect~\cite{SW},
which vanishes in the limit where the gravitational constant 
$G \to 0$  or the light velocity $c \to \infty$. Similarly to the 
displacement attained by swimmers in low Reynolds number 
fluids~\cite{C}-\cite{PB}, the displacement attained by swimmers 
in some given spacetime only depends on their local stroke. 

\begin{figure}[t]
\includegraphics[width=0.32\textwidth]{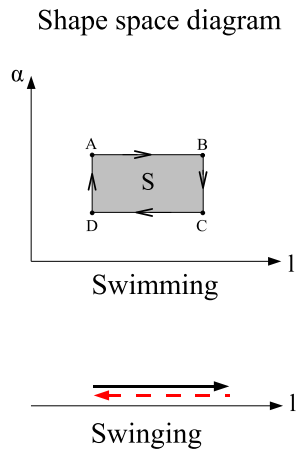}
\caption{The top figure carries the information that the 
swimming effect is a result of the non-zero area
of the square (in the shape space diagram) associated 
with the body deformation. Were the deformation
such that the area were null, the swimming
effect would vanish. The bottom figure illustrates
precisely this situation. Space displacements can be achieved 
in this case, however, through the swinging effect, i.e.
by expanding and contracting the legs at {\em different} 
points of the trajectory.}    
\label{shapespace}
\end{figure}

The fact that the swimming effect is purely relativistic
has caused some perplexity~\cite{La}-\cite{L}, 
since it has been known for 
a long time that there is a similar {\em classical effect} 
in non-uniform Newtonian gravitational fields, which is
present when $c \to \infty$. For example, an orbiting 
dumbbell-shaped body can modify its trajectory by contracting 
the strut connecting the two masses at one point and expanding 
it at another one~\cite{LM}. We stress here that this is a 
{\em nonlocal} effect, which 
appears due to the fact that the work performed by the dumbbell 
engine against the gravitational tidal force during the 
contraction differs from the one during the expansion. It 
is the resulting net work what allows the dumbbell to 
change from, say, a  bounded to an unbounded orbit 
(see Fig.~\ref{shapespace}). The shorter is the period of the whole 
contraction-expansion process, the smaller is the change of the 
trajectory, although this cannot be made arbitrarily small 
if one requires that the deformation velocity does not exceed 
$c$. This is in analogy with playground swings, where the 
oscillation amplitude is modified by an individual 
through standing and squatting in synchrony with the 
swing motion~\cite{Wa}. 
 
Here we perform a direct numerical simulation for a falling 
tripod to show that for fast enough cyclic deformations the 
swimming effect dominates over the swinging effect, 
while for slow enough cycles the opposite is true. 
We expect, thus, to set down any confusion concerning 
the independency of both effect. In addition, we calculate
and discuss the idiosyncratic features of a higher order 
term beyond the leading one obtained by Wisdom and extend 
the whole result to be applicable in cases where the tripod 
is in large red-shift regions.

\begin{figure}[t]
\includegraphics[width=0.2\textwidth]{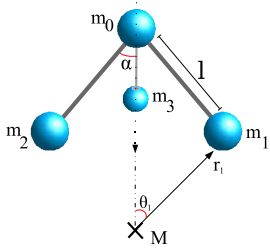}
\caption{The positions of the masses  $m_a$ $(a=0,1,2,3)$ 
are given through usual spherical coordinates 
$r_a, \theta_a, \phi_a$ with origin at the central mass $M$.} 
\label{tripodgeometry}
\end{figure}

\begin{figure}
\includegraphics[width=0.46\textwidth]{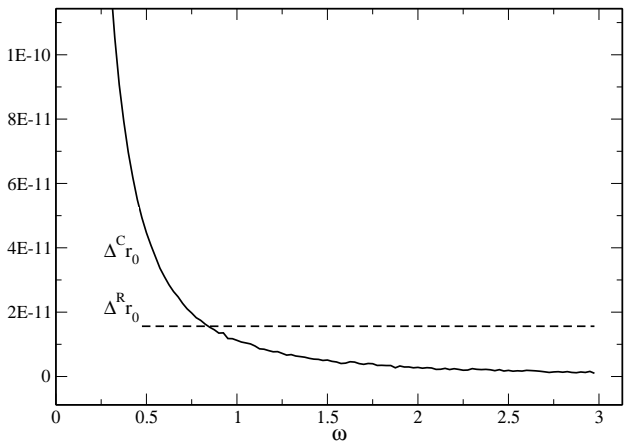}
\caption{ The full and dashed lines show 
$\Delta^{\rm C} r_0 \equiv 
          (r_0^{\rm quasi-rigid} - r_0^{\rm rigid})_{\rm clas}$
and 
$\Delta^{\rm R} r_0 \equiv 
          (r_0^{\rm quasi-rigid} - r_0^{\rm rigid})_{\rm rel}$, 
i.e. how much a free falling quasi-rigid tripod fails to follow 
a rigid one at the end of a complete cycle assuming
a Newtonian gravitational field and a Schwarzschild spacetime
characterized by a central mass $GM=1$, respectively.
Here the tripod is assumed to change its shape 
as shown in Fig.~\ref{snapshots} and $\omega \equiv 1/T$ 
is the cycle frequency. The rigid and quasi-rigid tripods are 
set free simultaneously with $G m_a = 0.1 $ and $r_0 =100$ 
($a=0,1,2,3$). Initially $\alpha=1$ and $l = 1$ 
and they vary as much as $\Delta \alpha= -0.01$ 
and $\Delta l =0.01$ along the cycle. Each quarter of 
the whole cycle takes as long as $T/4$. (Here $c=1$.)}
\label{keyfigure}
\end{figure}

Let us begin considering a tripod falling along the radial axis
in the Newtonian gravitational field of a spherically symmetric 
static body with mass $M$.  The three tripod endpoint masses 
$m_i$ ($i=1,2,3$) are connected to the mass $m_0$ at the vertex 
through {\em straight massless} struts with length $l$. The tripod 
is set with its vertex mass above the three endpoint masses and 
aligned symmetrically with the radial axis in order that the three 
struts make a common angle $\alpha$ with it 
(see Fig.~\ref{tripodgeometry}). The tripod legs are designed 
to {\em contract and expand}, $l=l(t)$, and {\em open and close}, 
$\alpha = \alpha(t)$, as much as $\Delta l$ and $\Delta \alpha$, 
respectively, along a complete cycle as ruled {\em a priori} 
by some internal 
engine. The Lagrangian used in the action $S=\int L \, dt$
to describe the falling tripod is
\begin{equation}
L=\sum_{a=0}^{3}\frac{GM m_a}{r_a} 
  +
  \sum_{a=0}^{3} \frac{m_a}{2} 
  \left( 
  \dot{r}_a^2 
  + r_a^2 \dot{\theta}_a^2
  + r_a^2 \sin^2 \theta_a \dot{\phi_a}^2
  \right)
\label{lagnewt}
\end{equation}
where $`` \; \dot{} \; " \equiv d/dt$. The positions
of the masses  $m_a$, $a=0,1,2,3$, are given through
usual spherical coordinates $r_a, \theta_a, \phi_a$ 
with origin at the central mass $M$. The tripod is not 
assumed to rotate, $\dot{\phi}_i=0$, and the underlying 
symmetry guaranties that $r_i=r_j$ and $\theta_i=\theta_j$ 
for $i,j=1,2,3$. The evolution $r_0 = r_0(t)$
($\theta_0=\phi_0=0$) of $m_0$ is given by numerically 
integrating the corresponding Euler-Lagrange equations
with the constraints 
\begin{equation}
r_i = (r_0^2 + l^2 -2r_0l \cos \alpha)^{1/2},\;
\theta_i = \arcsin \left( (l/r_i ) \sin \alpha \right)
\label{constr}
\end{equation} 
and $\phi_i = 2 \pi (i-1)/3 $. 
Here we consider $r_0$ and $p_{r_0}$ as the only 
independent dynamical variables. ($r_i$ and $\theta_i$ are
implicit functions of $r_0$ through Eq.~({\ref{constr}}).)
The solid line in 
Fig.~\ref{keyfigure} shows how much a quasi-rigid 
tripod changing shape as shown in Fig.~\ref{snapshots} 
fails to follow a rigid one at the end of a 
complete cycle, where both tripods are 
let free simultaneously and we have assumed that 
each quarter of the whole cycle takes as long as
$T/4$ of the total period $T$. 
(For the sake of comparison we use the position of $m_0$.) 
Clearly the slower (faster) 
is the cycle, the larger (smaller) is $\Delta^{\rm C} r_0$.

Let us examine in detail the high-frequency shape deformation 
region, $\omega \equiv 1/T \gg \sqrt{GM/(r_0^2 \Delta l)}$, 
for the sake of further comparison with the swimming effect. 
By ``high-frequency" we mean that along the whole period $T$
the tripods do not fall much in comparison with $\Delta l$.
We shall assume in this regime that $p_{r_0}$ is arbitrarily
small and approximately conserved: 
$p_{r_0}  =\partial L/\partial\dot{r_0}  \approx 0$.
As a result one  obtains, for 
$m_i = m_j$, $i,j=1,2,3$, 
\begin{equation}
dr_0 \approx U \,dl + V \,d\alpha ,
\label{drgeral}
\end{equation}
where $dr_0 = \dot r_0 dt$ and
$$
U =-
\frac{
      ({\partial r_1}/{\partial r_0})\,
      ({\partial r_1}/{\partial l})
      + r_1^2
      ({\partial \theta_1}/{\partial r_0})\,
      ({\partial \theta_1}/{\partial l})
     }
{m_0/(3m_1) +  
({\partial r_1}/{\partial r_0})^2 
+ r_1^2({\partial \theta_1}/{\partial r_0})^2} 
$$
and
$$
V =-
\frac{
({\partial r_1}/{\partial r_0})\,
      ({\partial r_1}/{\partial \alpha})
      + r_1^2
      ({\partial \theta_1}/{\partial r_0})\,
      ({\partial \theta_1}/{\partial \alpha})
     }
{m_0/(3m_1) +  
({\partial r_1}/{\partial r_0})^2 
+ r_1^2({\partial \theta_1}/{\partial r_0})^2} .
$$ 
The net translation accomplished after the complete cycle
ABCDA shown in Fig.~\ref{shapespace}, which circumvents 
an area $S$, can be computed using the Stokes theorem 
\begin{equation}
\Delta^{\rm C} r_0 \approx \int_{\partial S} 
          ( {\partial V}/{\partial l} -
            {\partial U}/{\partial \alpha}
          ) \;
          dl \wedge d\alpha,
\label{deltargeral}
\end{equation}
where $dl$ and $ d\alpha$ are treated as one-forms in the
shape space manifold covered with coordinates 
$\{ l, \alpha\}$. Now, because 
$\partial U/\partial\alpha = \partial V/\partial l$
[see Eq.~(\ref{constr})], 
we have that in this regime 
$\Delta^{\rm C} r_0 \approx 0$. 
Indeed, by associating the gravitational potential energy 
gained by the tripod along the process with the work performed 
against the gravitational tidal forces, we can estimate that 
\begin{equation}
\Delta^{\rm C} r_0 \approx {a G M l \Delta l}/{(r_0^4 \omega^2)}
\label{swingapprox}
\end{equation}
where $a$ is a constant, which depends on the detailed geometry
of the body. For the parameters chosen in Fig.~\ref{keyfigure}
$a \approx 0.1$. The fact that 
$\Delta^{\rm C} r_0 \stackrel{\omega \to \infty}{\longrightarrow} 0$ 
is a general result because in the high-frequency regime 
the one-form  $d r_0$ will be approximately closed
for any classical (or, even, semi-classical) 
potentials with no velocity dependence. 
Free-falling panicking individuals performing 
fast cyclic motions in Newtonian-like gravitational 
fields will not be able to change significantly 
their trajectories despite  the strength of their 
local stroke; it had better that they swing suitably
with low frequencies.   

Next, let us investigate how the above picture is modified when
one replaces the Newtonian gravitational field by the 
curved Schwarzschild spacetime associated with a spherically 
symmetric body with mass $M$ as described by the line element 
\begin{equation}
ds^2 =  f(r) c^2 dt^2  
        -f(r)^{-1} dr^2 
        - r^2 (d\theta^2 + \sin^2 \theta d\phi^2) ,
\label{Schwarzschild}
\end{equation}
where $f(r) = 1-2GM/c^2r$. The Lagrangian used in the
action $S= \int L \, dt$ to evolve 
the tripod is
\begin{equation}
L = \sum_{a=0}^3 m_a 
    \left(c^2 f_a  
          - \dot{r_a}^2 f_a^{-1}  
          - r_a^2 \dot{\theta_a}^2 
          - r_a^2 \sin^2 \theta_a \, \dot{\phi_a}^2
    \right)^{1/2}
\label{lagrg}
\end{equation}
where 
$f_a \equiv f(r_a)$ and 
$`` \; \dot{} \; " \equiv d/dt$. 
The constraints 
$ r_1= r_1(r_0,l,\alpha)$ 
and 
$\theta_1= \theta_1(r_0,l,\alpha)$ 
are obtained in this case by requiring that the 
tripod struts be geodesics in the $t \approx {\rm  const}$ 
space section of the static observers (with 4-velocity 
$u \propto \partial/ \partial t$), who measure 
$l=l(t)$ as the struts' proper length and
$\alpha = \alpha(t)$ as the proper angle of 
the struts with the radial axis.
(The ``$\approx $" used above is because 
although we are in the high-frequency regime, 
it takes some time to complete each cycle.) 
Now, it is convenient to expand 
Eq.~(\ref{lagrg}) up to order $v^{2}/c^{2}$ to avoid nonlinear 
equations. We obtain, then, in the high-frequency regime,
$\omega \equiv 1/T\gg \sqrt{GMf_0^{1/2}/( r_0^2 \, \Delta l) }$
with $T$ being the total {\em coordinate} period,
the relativistic analogue of Eq.~(\ref{drgeral}):
\begin{equation}
dr_0 \approx X dl+ Y d\alpha , 
\label{rdotrg}
\end{equation}
where 
$$
X \! = \!\!
 \frac{-
      ({\partial r_1}/{\partial r_0})
      ({\partial r_1}/{\partial l})
       -
      f_1 r_1^2 
      ({\partial \theta_1}/{\partial r_0})
      ({\partial \theta_1}/{\partial l}
      )
     }{(m_0/3 m_1) (f_1/f_0)^{3/2} 
      + 
 ({\partial r_{1}}/{\partial r_{0}})^2 
+ f_1 r_1^2 ({\partial \theta_{1}}/{\partial r_{0}})^{2}}
$$
and
\begin{equation*}
Y \!\! = \!\!
 \frac{-
      ({\partial r_1}/{\partial r_0})
      ({\partial r_1}/{\partial \alpha})
       -
      f_1 r_1^2 
      ({\partial \theta_1}/{\partial r_0})
      ({\partial \theta_1}/{\partial \alpha}
      )
     }{(m_0/3 m_1) (f_1/f_0)^{3/2} 
      + 
 ({\partial r_{1}}/{\partial r_{0}})^2 
+ f_1 r_1^2 ({\partial \theta_{1}}/{\partial r_{0}})^{2}}.
\end{equation*}
Afterwards, we  integrate Eq.~(\ref{rdotrg}), 
\begin{equation}
\Delta^{\rm R} r_0 \approx \int_{\partial S} 
          ( {\partial Y}/{\partial l} -
            {\partial X}/{\partial \alpha}
          ) \;
          dl \wedge d\alpha ,
\label{deltargeral2}
\end{equation}
along the complete cycle ABCDA (see Fig.~\ref{shapespace}), 
obtaining for small enough $l / r_0$ and 
$\Delta \alpha$, $\Delta l$ 
\begin{widetext}
\begin{equation}
\Delta^{\rm R} r_0 \! \approx \!
       \frac{-3m_0 m_1 }
            {(m_0 + 3m_1 )^2 }
       \frac{GM}{c^2 r_0} \sqrt{f_0}
          \left[ 
          \frac{l^2}{r_0^2} 
          + 
          \left( 
              \frac{3 m_1}{m_0} \sqrt{f_0}
            + \frac{m_0 - 3m_1}{m_0 + 3m_1} 
              \frac{GM}{c^2 r_0 \sqrt{f_0}} 
          \right) 
          \frac{l^3}{r_0^3} \cos \alpha   
          \right] 
         \sin \alpha \Delta \alpha \Delta l  
\label{swimfinal}
\end{equation}
\end{widetext}
which corresponds to a proper distance 
$ \Delta \lambda \approx \Delta^{\rm R} r_0/ \sqrt{f_0} $
as measured by the static observers assuming 
$ \Delta^{\rm R} r_0/r_0 \ll 1 $. A numerical integration
of Eq.~(\ref{rdotrg}) with no restriction on 
$\Delta \alpha$ and $\Delta l$ was performed and is in 
agreement with Eq.~(\ref{swimfinal}) in the proper limit.  
Assuming that
the leading term in this equation dominates over
the next order one,
we conclude that $\Delta \lambda \ll \Delta l$.
The term of order $l^2/r_0^2$ in Eq.~(\ref{swimfinal}) coincides 
with the result obtained in Ref.~\cite{W} for $r_0 \gg 2 GM/c^2$ 
and goes beyond, since it also holds close to the horizon: 
$r_0 $
{\raise0.3ex\hbox{$\;>$\kern-0.75em\raise-1.1ex\hbox{$\sim\;$}}}
$2 GM/c^2$. It is interesting to note that the leading term
of $\Delta^{\rm R} r_0$ tends to decrease as the tripod approaches 
the horizon. This can be understood
from the fact that assuming that $l$ is fixed, the coordinate 
size  of the tripod decreases as $l \sqrt{f_0}$. 
As a result, the tripod is only able to probe smaller 
{\em coordinate} size regions. Now, close to the horizon
the $t-r$ section of the
Schwarzschild line element~(\ref{Schwarzschild}) 
can be approximated ($\theta, \phi \approx {\rm const}$) 
by the Rindler wedge one~\cite{Ri}: 
$
ds^2 \approx (\rho c^2/4GM)^2 c^2 dt^2 - d\rho^2 ,
$ 
where 
$\rho = (4GM/c^2)/ \sqrt{f(r)^{-1}-1}$, 
which has vanishing curvature. 
Thus, for the same reason $\Delta^{\rm R} r_0$
{\em vanishes} in {\em flat} spacetimes, this is 
{\em damped} in the {\em horizon's neighborhood}. 
Clearly, it remains the fact that the corresponding
$\Delta \lambda$ not only is not damped but increases as 
the tripod approaches the horizon, as a consequence of 
the fact that the space curvature gets larger. Concerning 
the next order term, it is interesting
to note that it can be positive, negative or null depending
on the masses and tripod position. 
$\Delta^{\rm R} r_0$ is plotted 
in Fig.~\ref{keyfigure} 
(see dashed line) as a constant in the {\em  high-frequency} 
region. We have taken care to keep the deformation
velocity $v < c$. For the frequency range shown in the 
Fig.~\ref{keyfigure}, we have 
$10^{-2}$
{\raise0.3ex\hbox{$\;<$\kern-0.75em\raise-1.1ex\hbox{$\sim\;$}}}
$v/c$
{\raise0.3ex\hbox{$\;<$\kern-0.75em\raise-1.1ex\hbox{$\sim\;$}}}
$10^{-1}$.
We see that for high enough frequencies the swimming effect
can dominate the swinging effect by orders of magnitude. 
For the parameters chosen in the graph, the swimming effect
begins to dominate over the swinging effect at 
$\omega$
{\raise0.3ex\hbox{$\;>$\kern-0.75em\raise-1.1ex\hbox{$\sim\;$}}}
$0.9$. This can be estimated analytically quite well  by equating 
Eqs.~(\ref{swingapprox}) and~(\ref{swimfinal}). A full 
general-relativistic numerical simulation, which would involve 
formidable difficulties associated with the relativistic 
rigid body concept, is expected
to approach smoothly the swinging and swimming predictions 
in the low- and high-frequency regions, respectively. 
(For a movie on the swinging and swimming effects see 
Ref.~\cite{movie}.)

It seems to be a challenging problem to take into 
account the decrease of the quasi-rigid body mass 
(i.e, rest energy) as a consequence 
of the swimming. This is desirable when the work
$W$ spent (or gained) along the process is of order of
$(m_0 + 3 m_1) c^2$. The work associated with a 
displacement $\Delta^{\rm R} r_0 $ 
can be estimated for $W \ll (m_0 + 3 m_1) c^2$
to be 
\begin{equation}
W\approx (m_0 + 3m_1) 
GM \Delta^R r_0 /(f_0 r_0^2 ),
\label{work}
\end{equation}
where 
${\Delta^R r_0}/ r_0  \ll f_0 r_0 c^2 /(GM) $. 
Eq.~(\ref{work}) suggests that this is very 
costly to swim close to the horizon. Actually, 
even far away from it, we do not expect the tripod to 
be able to climb upwards the space. This can be seen 
as follows. Along a complete period $T$,
the free rigid tripod falls down about         
$
\Delta^{\rm F} r_0 \approx {GM f_0 T^2}/{(2 r_0^2)} 
$.
By imposing that the deformation velocity  
$v$
{\raise0.3ex\hbox{$\;<$\kern-0.75em\raise-1.1ex\hbox{$\sim\;$}}}
$c$, 
we obtain 
$T$
{\raise0.3ex\hbox{$\;>$\kern-0.75em\raise-1.1ex\hbox{$\sim\;$}}}
$l / (c \sqrt{f_0}) $
and, thus,
$
\Delta^{\rm F} r_0 
$
{\raise0.3ex\hbox{$\;>$\kern-0.75em\raise-1.1ex\hbox{$\sim\;$}}}
$
{GM l^2}/{2 c^2 r_0^2 } > \Delta^{\rm R} r_0
$.
This raises the interesting ``engineering" issue 
concerning what would be the most efficient geometry
and stroke for quasi-rigid spacetime swimming bodies. 
In this vein, it would be also interesting to see how the tripod 
could accomplish more complex maneuvers through
asymmetric deformations. 
This is remarkable that General Relativity, which is a 
quite studied ninety-years-old theory did not loose 
its gift of surprising us. After all, free-falling 
panicking individuals may change  their trajectories 
by doing fast cyclic motions because the world is 
relativistic. 
  
EG is indebted to J. Wisdom for various conversations and
P. Letelier for the support. 
Two of us, CM and GM, would like to acknowledge full and partial
financial supports from Funda\c c\~ao de Amparo \`a Pesquisa do 
Estado de S\~ao Paulo, respectively, while GM is also thankful 
to Conselho Nacional de Desenvolvimento Cient\'\i fico e 
Tecnol\'ogico  for partial support.

\end{document}